\documentclass[aps,prl,pdflatex,reprint,showpacs,superscriptaddress]{revtex4-2}
\usepackage{amsmath}
\usepackage{amssymb}
\usepackage{amsthm}
\usepackage{bbm}
\usepackage{graphicx}
\usepackage{hyperref}
\usepackage{physics}
\usepackage{systeme}
\usepackage{times}
\usepackage{epstopdf}
\usepackage{float}
\usepackage{xcolor}
\usepackage[normalem]{ulem}

\hypersetup{
  colorlinks=true,linkcolor=blue,citecolor=blue,
  filecolor=blue,urlcolor=blue,breaklinks=true
}

\begin{document}
\title{Theoretical model for the description of a single quantum dot using geometry}
\author{Francisco A. G. de Lira}
\email{francisco.augoncalves@gmail.com}
\affiliation{
        Departamento de F\'{i}sica,
        Universidade Federal do Maranh\~{a}o,
        65085-580, S\~{a}o Lu\'{i}s, Maranh\~{a}o, Brazil
      }

\author{Edilberto O. Silva}
\email{edilberto.silva@ufma.br}
\affiliation{
Departamento de F\'{i}sica,
Universidade Federal do Maranh\~{a}o,
65085-580, S\~{a}o Lu\'{i}s, Maranh\~{a}o, Brazil
}
\date{\today }

\begin{abstract}
An alternative model to describe the electronic and thermal properties of quantum dot based on triangle geometry is proposed. The model predicts characteristics and limitations of the system by controlling the magnetic field and temperature, and the other parameters involved. 
\end{abstract}

\pacs{68.65.-k, 68.65.Hb, 73.21.-b, 73.21.La,73.22.-f}
\maketitle

Mesoscopic systems are among the most important and successful models describing physical reality. They are represented by geometrical structures, where the states of the system are  described by points, the observables are real-valued functions, and the quantum evolution is governed by a Hamiltonian operator \cite{harrison2016quantum,joyce2006quantum,RMP.2002.74.1283,RMP.2007.79.1217,RMP.2000.72.895}. Semiconductor structures such as InAs and GaAs (among other semiconductors) have been studied extensively in recent years and continue to be objects of study because of their important physical properties and their great potential for applications in technology \cite{PE.2020.118.113913,PRL.1996.76.3005,PE.2021.134.114904,PRB.2021.104.195411,CBM.2012.9.151,JL.2019.205.287,ML.2020.268.127595,PRB.2021.104.195306,PLA.2005.336.434,SM.2021.156.106919,PRL.1990.64.2559,MSSP.2021.124.105614,PRB.1998.57.7190,PRB.1992.46.9780,OC.2010.283.1510,OC.2013.287.241,PE.2019114.113629,PLA.2012.376.2712}. It is well known that geometry is present in physics at all scales of measures. The  possibility of associating the expression of a physical quantity to some simple geometric shape would allow us to know several characteristics of the system in a more direct and simplified way. In this letter, we propose a model that maps the energy spectrum of a quantum dot to formula of the area of an  isosceles triangle. From this result, other system properties can be accessed in a simple way. We use the GaAs quantum dot as an application of the model.

\begin{figure}[!h]
	\centering
	\includegraphics[scale=0.5]{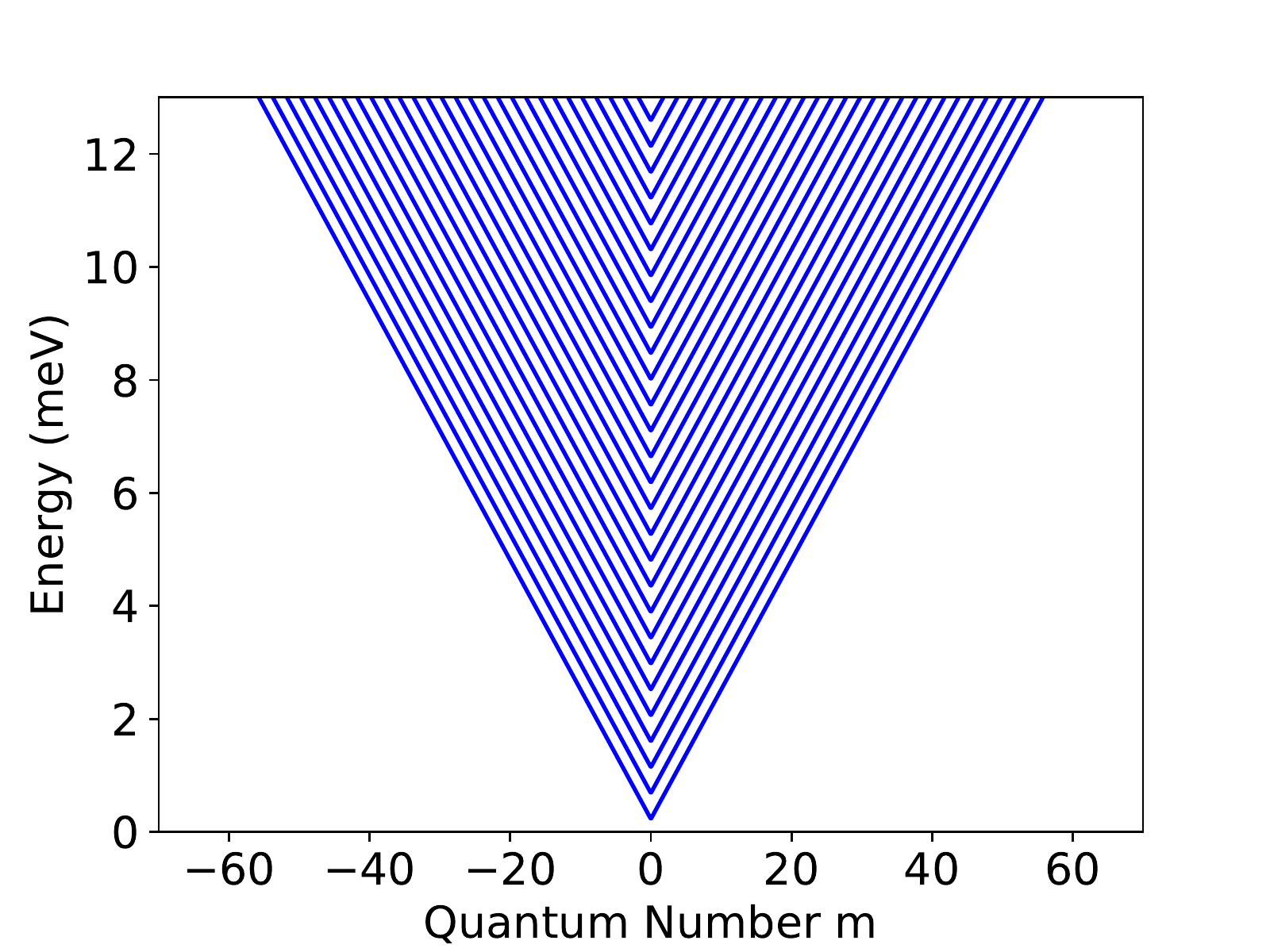}
    \caption{The first twenty-seven energy lowest subbands for a quantum dot with $\hbar\omega_{0}=0.458\hspace{0.05cm}\text{meV}$ in zero magnetic field.}
	\label{energym}		
\end{figure}

The expression for the energy eigenvalues of an electron at a parabolic confining potential $V(r)=\frac{1}{2}\mu_{e}\Omega_{0}^{2}r^{2}$ ($\Omega_{0}$ is the confinement potential
strength and $\mu_{e}$ is the effective mass of the electrons) in the presence of an external uniform magnetic field $B$ perpendicular to the surface is known to be  \cite{PRB.2006.73.155315}
\begin{equation}
E_{n,m}=\left(2n+|m|+1\right)\hbar\Omega-\dfrac{1}{2}m\hbar\omega_{c},\label{aqd03}
\end{equation}
where $n=0,1,2,\dots$ and $m=0,\pm1,\pm2,\dots$are the principal and azimutal quantum numbers, respectively; $\Omega=\sqrt{\Omega_{0}^{2}+\omega^{2}_{c}/4}$ represents the effective frequency and $\omega_{c}=eB/\mu_{e}$ is the cyclotron frequency. In order to compare with previous results in the literature, we redefine $\Omega=\omega/2$, with $\omega=\sqrt{\omega^{2}_{0}+\omega^{2}_{c}}$ and $\omega_{0}=2\Omega_{0}$. For a quantum dot in the absence of magnetic field, we get \cite{PRB.1999.60.5626}
\begin{equation}
E_{n,m}=\left(n+\dfrac{1}{2}+\dfrac{\abs{m}}{2}\right)\hbar\omega_{0}.
\label{aqd06}
\end{equation}
Equation (\ref{aqd06}) allow us to identify that electrons are grouped in families of states with the same subband index $n$ (Fig. \ref{energym} with $\hbar\omega_{0}=0.458\hspace{0.05cm}\text{meV}$).
These subbands are symmetric about the energies with $m=0$, where each electron of the system occupies a given state labelled by $n$ and $m$ quantum numbers. Then, electrons must fill partially the subbands until they establish themselves in the lowest energy configuration. 

\begin{figure}[!h]
\centering
\includegraphics[scale=0.56]{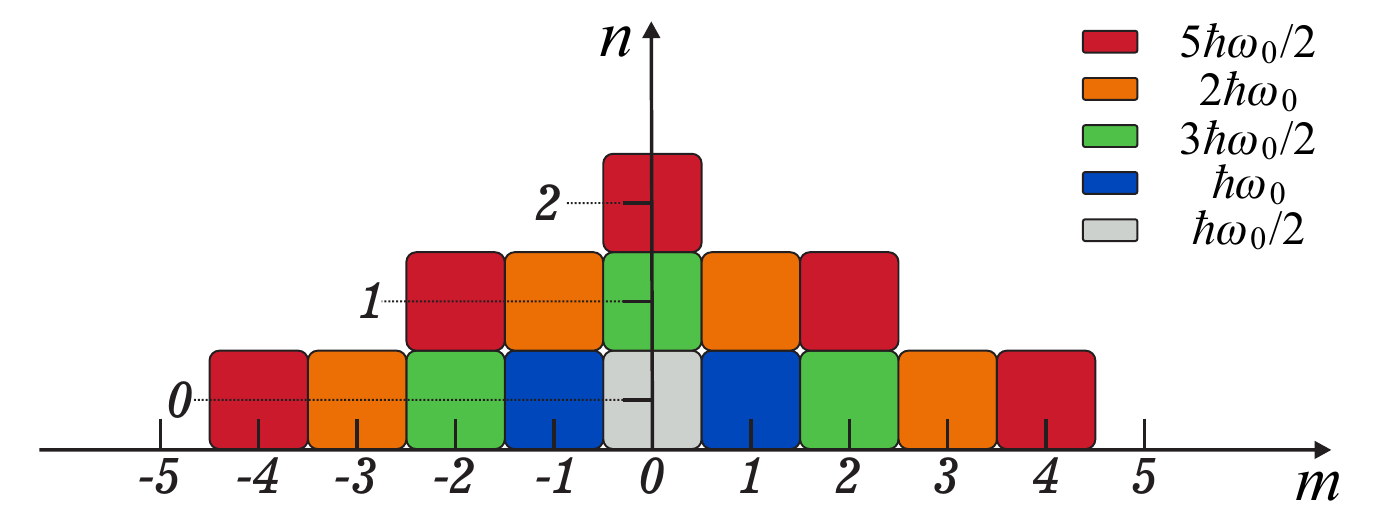}
\caption{Illustration of the energy levels in the $n \times m$ state space. Each block represents a given $\psi_{n,m}$ state and localize two electrons with opposite spins. A given block color represents specific energy levels which are shifted by multiples of $\hbar\omega_{0}/2$.}
\label{degerency}	
\end{figure}

\begin{figure}[!h]
\centering
\includegraphics[scale=0.5]{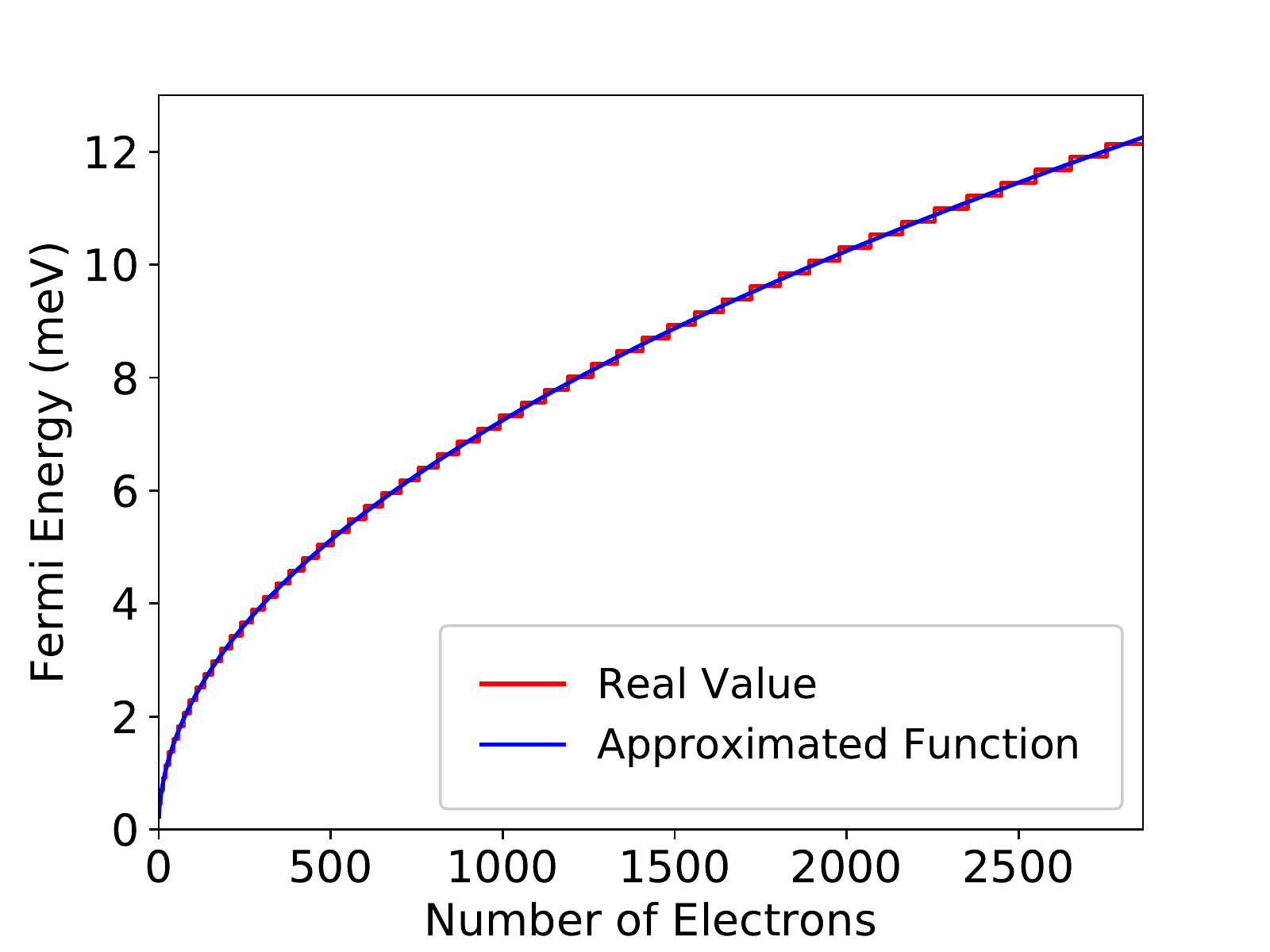}
\caption{Comparison between the exact Fermi energy values and those provided by approximated function $\epsilon_{f}$. For high values of $N_{e}$, $\epsilon_{f}$ provides a more accurate description because the steps become wider.}
\label{comparison}	
\end{figure}

We can find the density of states and hence the Fermi energy by defining each state as a unitary area block in a $n\times m$ space of state and populate them with a few electrons in pairs one by one from the lowest energetic state to the highest allowed one. This is illustrated in the diagram of Fig. \ref{degerency}.
We can see that degeneracy grows linearly with the energy level (color blocks), where the eigenvalue of the highest occupied one is called Fermi energy $\epsilon_{f}$. For a sample containing an even number of electrons denoted by $N_{e}$, we have $N_{e}/2$ blocks fully occupied in the diagram. By fixing $E_{n,m}=\epsilon_{f}$, Eq. (\ref{aqd06}) provides
\begin{equation}
S=\dfrac{\epsilon_{f}}{\hbar\omega_{0}}-\dfrac{1}{2}=n+\dfrac{\abs{m}}{2}.
\label{aqd07}
\end{equation}
Equation (\ref{aqd07}) describes a ``{\it Fermi curve}" which bounds the highest populated energy states from the unoccupied region in the $n\times m$ space. If we are capable of calculating the area under this curve in terms of $S$, then we will be able to estimate the Fermi energy as a function of $N_{e}$ because the total area of diagram must be equal to $N_{e}/2$.
\begin{figure}[H]
\centering
\vspace{-0.25cm}
\includegraphics[scale=0.45]{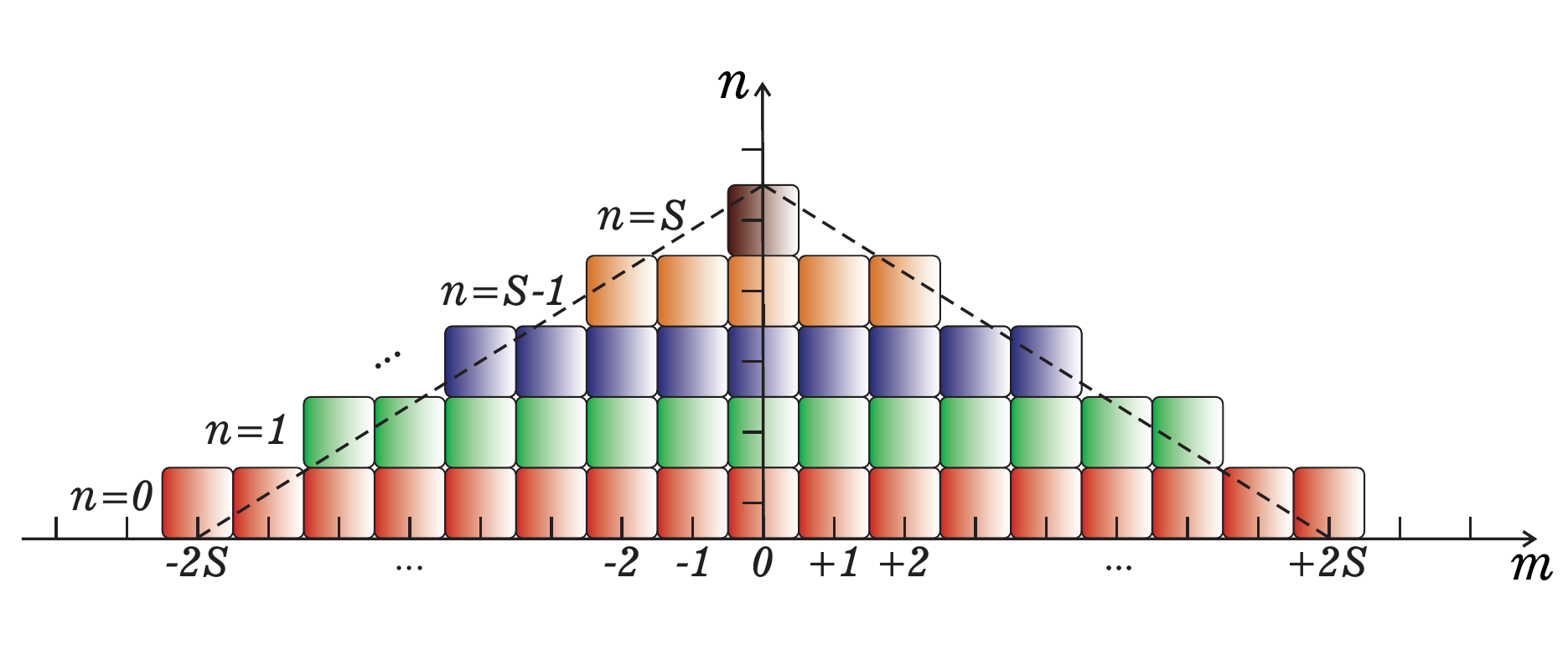}
\caption{Diagram representing the space of state and its respective area $N_{e}/2$. For $N_{e}>>1$, we can approximate the area of the triangle $4S(S+1)/2$ defined by the dashed line to the exact area of the diagram $N_{e}/2$.}
\label{diagrama}	
\end{figure}
\noindent In order for this to be accomplished, we consider $N_{e}>>1$. In this case, the total area of the diagram can be approximated to the area of the triangle defined by the dashed line, whose value is $2S(S+1)$ (Fig. \ref{diagrama}). Thus, the corresponding solution for the Fermi energy is
\begin{equation}
\epsilon_{f}=\dfrac{\sqrt{1+N_{e}}}{2}\hbar\omega_{0}.
\label{aqd09}
\end{equation}
For a sample containing $2800$ electrons, with $\mu_{e}=0.067\,m_{e}$, we have $\epsilon_{f}\approx 12.12$ meV. This value is in good agreement with the numerical result $\epsilon_{f} \approx 12.2$ meV of Ref. \cite{PRB.1999.60.5626}. This is most evident when we compare the exact values of the Fermi energy for different values of $N_{e}$ and the behavior provided by the function (\ref{aqd09}) of which we can observe that the function $\epsilon_{f}$ gives a more accurate description for a large $N_{e}$ (Fig. \ref{comparison}).
Equation (\ref{aqd09}) also determines the density of states per unity of energy,
\begin{equation}
D(\epsilon)=\dfrac{d N_{e}}{d\epsilon}=\dfrac{8\epsilon}{(\hbar\omega_{0})^{2}},
\label{aqd10}
\end{equation}

\noindent where, as discussed in Fig. \ref{degerency}, it is expected this density to increase linearly in energy, because after each shift of $\hbar\omega_{0}/2$, the number of states with opposite spins in the energy level increases by two. The energy variable $\epsilon$ represents a bounding value which encloses all the occupied states with energy lower than $\epsilon$. The highest value for $\epsilon$ in absolute zero is $\epsilon_{f}$.

Using (\ref{aqd10}), we can find an expression for both internal energy and pressure in absolute zero. From the relation $U_{0}=\int_{0}^{\epsilon_{f}}D(\epsilon)\epsilon d\epsilon$, we obtain
\begin{equation}
U_{0}\approx \dfrac{2N_{e}}{3}\epsilon_{f},
\label{aqd11}
\end{equation}
which gives a correspondent value $U_{0} \approx 22.6\,\mu$eV, very similar to what we should have obtained if the eigenvalues of all occupied states were computed individually.

At this stage, we argue that due to the Pauli exclusion principle, it is natural for an internal force of quantum nature to appear, preventing the system from a collapse when it is cooled down. Using the relation between the width of the quantum dot and the Fermi energy \cite{PRB.1999.60.5626} expressed by $\Delta r=\sqrt{8\epsilon_{f}/(\mu_{e}\omega^{2}_{0})}$, we are able to relate the internal energy with the dot area $A$ through the relation $U_{0}\approx (4/3)\pi\hbar^{2}N^{2}_{e}/(\mu_{e}A) $. Differentiating it with respect to the area $A$ and defining the quantity $\sigma=N_{e}/A$ as the surface density of electrons, we obtain an expression of the form $dU_{0}=-P_{0}dA$ with
\begin{equation}
P_{0}=\dfrac{4}{3}\dfrac{\pi\hbar^{2}\sigma^{2}}{\mu_{e}}
\label{aqd16}
\end{equation}
representing an internal force per unit length that acts on the circumference of the dot against an external work compressing the dot surface by $dA$.

Despite expressed by meV/nm$^{2}$, which is a usual unit of surface tension, the phenomenology of the quantity $P_{0}$ correspond to a analogue of pressure for a two-dimensional system due to the negative sign in $dU_{0}$. If mechanical work is done on the sample, $\Delta r$ is reduced, and consequently leading to an increase in both $U_{0}$ and $\sigma$. Then, the system responds with higher values of $P_{0}$ in order to avoid its compression. On the other side, if the source of external work disappear suddenly, the system does work by expanding itself to the original configuration, increasing its area and reducing $\sigma$ and $U_{0}$. This behavior is pretty characteristic of a pressure.

Let us now consider the influence of thermal effects on the system. 
From Eq. (\ref{aqd10}), We can find an expression that denotes the state density per unit energy per unit area $D(\epsilon)/A=\mu_{e}/\pi\hbar^{2}$. Using this result, we can construct the integral $\gamma=\int_{0}^{\infty}D(\epsilon)f(\epsilon)/A d\epsilon$, where $f(\epsilon)$ is the Fermi-Dirac distribution function. Note that we can not assume $\gamma=\sigma$ because $A$ explicitly depends on the energy as $8\pi\epsilon/(\mu_{e}\omega^{2}_{0})$. The integral $\gamma$ can be solved for $\mu(T)$ to get
\begin{equation}
\mu(T)=k_{B}T \ln\Bigg[\exp(\dfrac{2\pi\hbar^{2}\sigma}{\mu k_{B}T})-1\Bigg],
\label{aqd21}
\end{equation}
from which it was determined $\gamma=2\sigma$ because $\mu(0)=\epsilon_{f}\approx 2\pi\hbar^{2}N_{e}/(\mu A)$. At temperatures $T$ much lower than the Fermi temperature $T_{f}=140.6$ K, the chemical potential becomes $\mu(T)\approx \epsilon_{f}$.

At finite temperatures the variation of energy in the system is defined by $\Delta U=\int_{0}^{\infty}D(\epsilon)f(\epsilon)\epsilon d\epsilon-\int_{0}^{\epsilon_{f}}D(\epsilon)\epsilon d\epsilon$.
Since $\mu$ is constant and equal to $\epsilon_{f} $ with good agreement for $T\le 4.2\hspace{0.05cm}\text{K}$, we can use the identity $N_{e}=\int_{0}^{\infty}D(\epsilon)f(\epsilon) d\epsilon=\int_{0}^{\epsilon_{f}}D(\epsilon) d\epsilon$, from which we derive the following expression for the heat capacity due to the electrons \cite{Book.2004.kittel}:
\begin{equation}
C_{e}\approx D(\epsilon_{f})\int_{0}^{\infty}\dfrac{df}{dT}(\epsilon-\epsilon_{f}) d\epsilon.
\label{aqd23}
\end{equation}
 Assuming that $K_{B}T<<\epsilon_{f}$ and making the change of variables $(\epsilon-\epsilon_{f})/k_{B}T\rightarrow x$, we find
\begin{equation}
C_{e}=\dfrac{\pi^{2}}{3}D(\epsilon_{f})k^{2}_{B}T=\dfrac{2\pi^{2}}{3}(N_{e}+1)k_{B}\dfrac{T}{T_{f}},
\label{aqd24}
\end{equation}
where we have rewritten $D(\epsilon)$ as $2(N_{e}+1)/\epsilon$. Therefore, we can write both thermal energy $U(T)$ and thermal pressure $P(T)$ in terms of $U_{0}$ and $P_{0}$, respectively, as
\begin{equation}
U(T)\approx U_{0}\Bigg[1+\dfrac{\pi^{2}}{2}\left(\dfrac{T}{T_{f}}\right)^{2}\Bigg]
\label{aqd26}
\end{equation}
and
\begin{equation}
P(T)=P_{0}\Bigg[1+\frac{\pi^{2}}{2}\left(\frac{T}{T_{f}}\right)^{2}\Bigg].
\label{aqd27}
\end{equation}
Result (\ref{aqd27}) is quite interesting. It shows that the temperature affects the internal pressure in the dot, causing  it to expand its area. Such effect has a significant importance for technological applications involving quantum dots since the optical properties are connected to their sizes through the band gap level which is inversely proportional to the dot area. In fact, from the relation between $\Delta r$ and $\epsilon_{f}$ using result (\ref{aqd09}), we note the variation in $A$ affects the quantum of energy by $\hbar\omega_{0}=(4\pi\hbar^{2}/{\mu_{e}A})\sqrt{N_{e}+1}$ \cite{agrawal2013introduction,Materials.2010.3.2260,APL.2011.98.193105}.

\begin{figure}[!ht]
\centering
\includegraphics[scale=0.5]{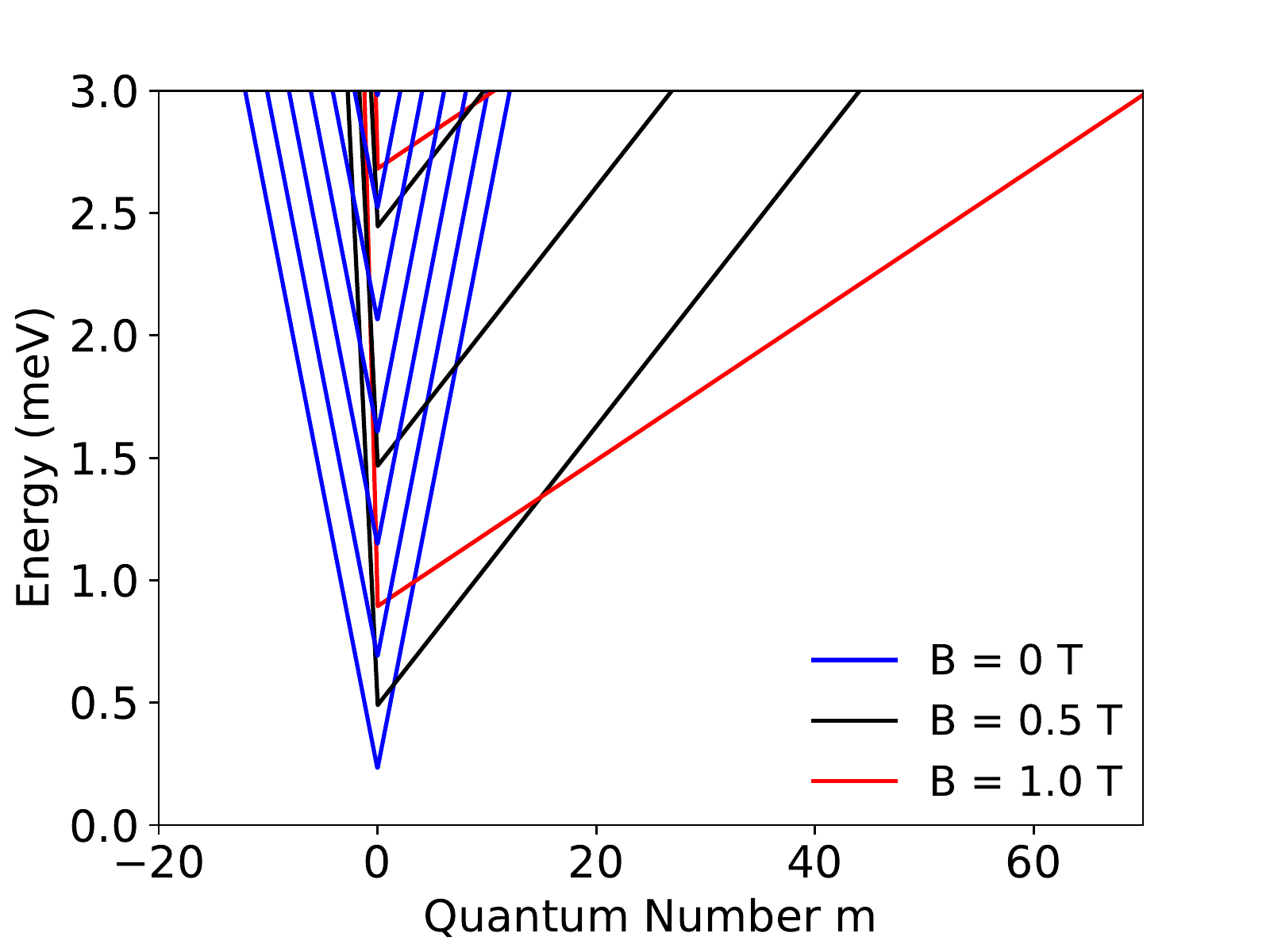}
\caption{Energies (Eq. (\ref{aqd270})) as a function of $m$ for $\hbar\omega_{0}=0.458$ meV. The energy gap between the subbands minimum increases as $B$ grows.}
\label{subbandmag}	
\end{figure}

When the quantum dot is placed in a uniform magnetic field its energy eigenvalues 
\begin{equation}
E_{n,m}=\left(n+\dfrac{1}{2}+\dfrac{\abs{m}}{2}\right)\hbar\omega-\dfrac{m}{2}\hbar\omega_{c}
\label{aqd270}
\end{equation}
show us that the shape, orientation and the position of the sub-bands are affected (Fig. \ref{subbandmag}).
When $B$ increases, both the states with $m>0$ and $m<0$ from the highest sub-bands become more energetic, and therefore, they are depopulated until the electrons accommodate themselves in the states with $m>0$ of the lowest sub-band.

\begin{figure}[!h]
\centering
\includegraphics[scale=0.5]{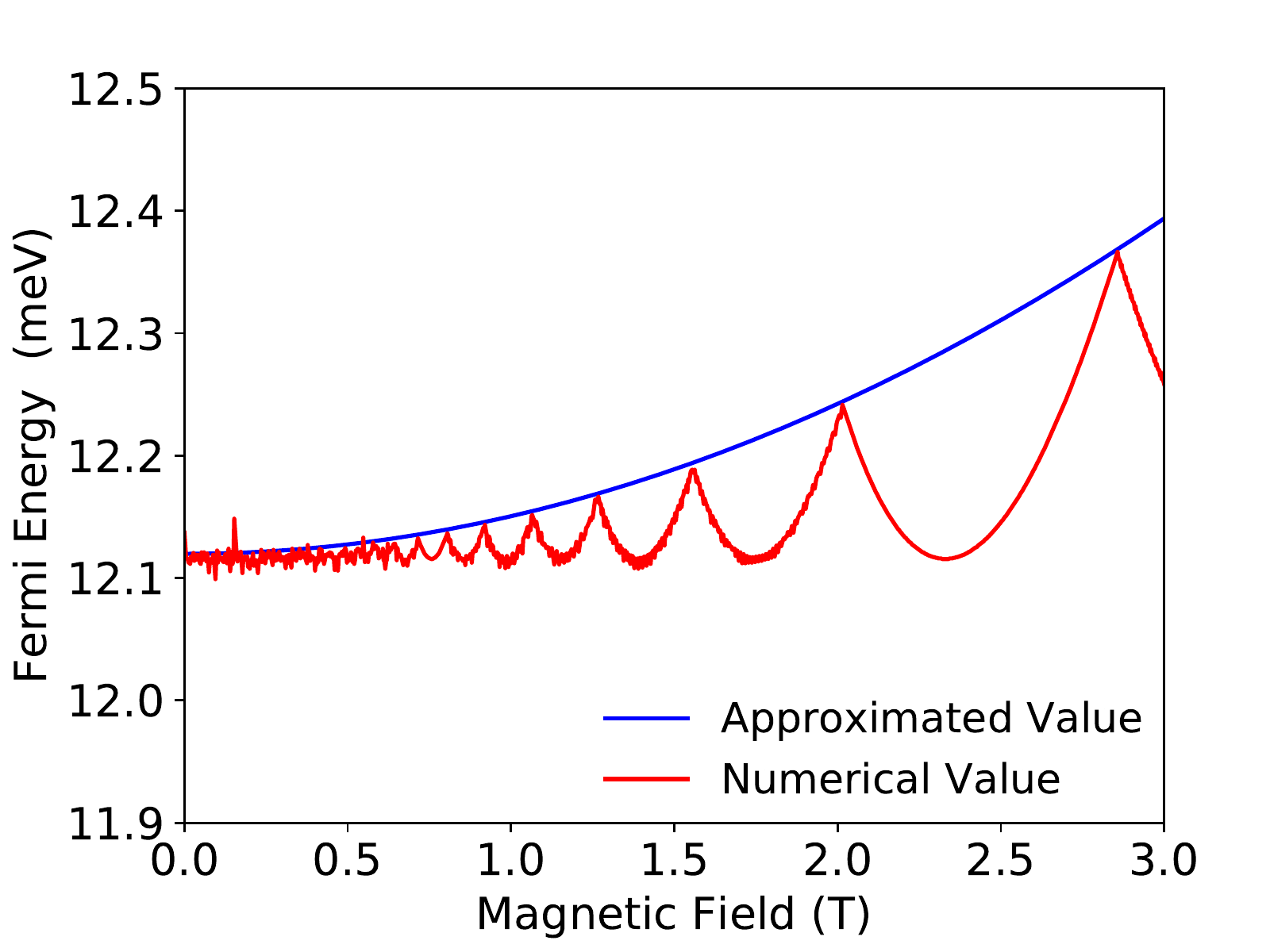}
\caption{Comparison between the approximated function for $\epsilon_{f}$ (Eq. \ref{aqd31}) and the numerical approach. The curve obtained from our model coincides with the peaks of the numerical curve. The points of maximum amplitudes indicate when a subband is completely depopulated.}
\label{fermimag}	
\end{figure}

Similar to the case free of magnetic field, we can determinate the total area of the block diagram and subsequently find an analytical expression for the Fermi energy $\epsilon_{f}$. Rewriting (\ref{aqd03}) as
 \begin{equation}
S=\dfrac{\epsilon_{f}}{\hbar\omega}-\dfrac{1}{2}=n+\dfrac{\abs{m}}{2}\left[1-\omega_{r}\text{sgn}(m)\right],
\label{aqd28}
\end{equation} 
where $\text{sgn}(m)$ is the signal function of $m$ and $\omega_{r}=\omega_{c}/\omega$, we find that
\begin{equation}
\dfrac{N_{e}}{2}=\dfrac{(4S)(S+1)}{2\left(1-\omega_{r}^{2}\right)},
\label{aqd30}
\end{equation}
which solved for $\epsilon_{f}$, gives
\begin{equation}
\epsilon_{f}=\frac{1}{2}\sqrt{\left(1-\omega_{r}^{2}\right)N_{e}+1}\,\hbar\omega.
\label{aqd31}
\end{equation}
Both numerical and $\epsilon_{f}$ description are showed in Fig. \ref{fermimag}. It is immediate to verify that for $\omega_{c}=0$, we recover the formula (\ref{aqd09}). On the other hand, for $\omega_{c}\rightarrow\infty$, we find $\epsilon_{f}=\hbar\omega_{c}/2$, which is exactly what we expect from Eq. (\ref{aqd270}) since in this regime all electrons must be in the lowest subband $(n = 0)$ with energy $\hbar\omega_{c}/2$.

From Eq. (\ref{aqd31}), we determine the density of states
\begin{equation}
D(\epsilon, B)=\dfrac{8\epsilon}{\left(1-\omega_{r}^{2} \right)(\hbar\omega)^{2}},
\label{aqd32}
\end{equation}
from which we can immediately see that for $\omega_{r}\rightarrow 1 $, we obtain $D(\epsilon)\rightarrow \infty$. In this regime, all occupied states have the same energy value $\hbar\omega_{c}/2$ as previously discussed.

The internal energy $U(B)$ at the absolute zero as a function of the magnetic field can be expressed, using the definition $ (\hbar\omega)^{2}=(\hbar\omega_{0})^{2}+(\hbar\omega_{c})^{2} $, as
\begin{equation}
U(B)=\sqrt{U^{2}_{0}+\frac{1}{9}N^{2}_{e}\hbar^{2}\omega_{c}^{2}},
\label{aqd34}
\end{equation}
where we clearly see that only the second term in the square root explicitly depends on the magnetic field $B$. From this result, we can determine the magnetization of the dot as a function of
magnetic-field strength by means of the formula $\mathcal{M}=-\partial U(B)/\partial B$. We obtain
\begin{equation}
\mathcal{M}(B)=-\frac{e\hbar N^{2}_{e}}{9\mu}\left( U^{2}_{0}+\dfrac{1}{9}N^{2}_{e}\hbar^{2}\omega_{c}^{2}\right)^{-\frac{1}{2}}\hbar\omega_{c}.
\label{aqd35}
\end{equation}

\begin{figure}[!ht]
\centering
\includegraphics[scale=0.5]{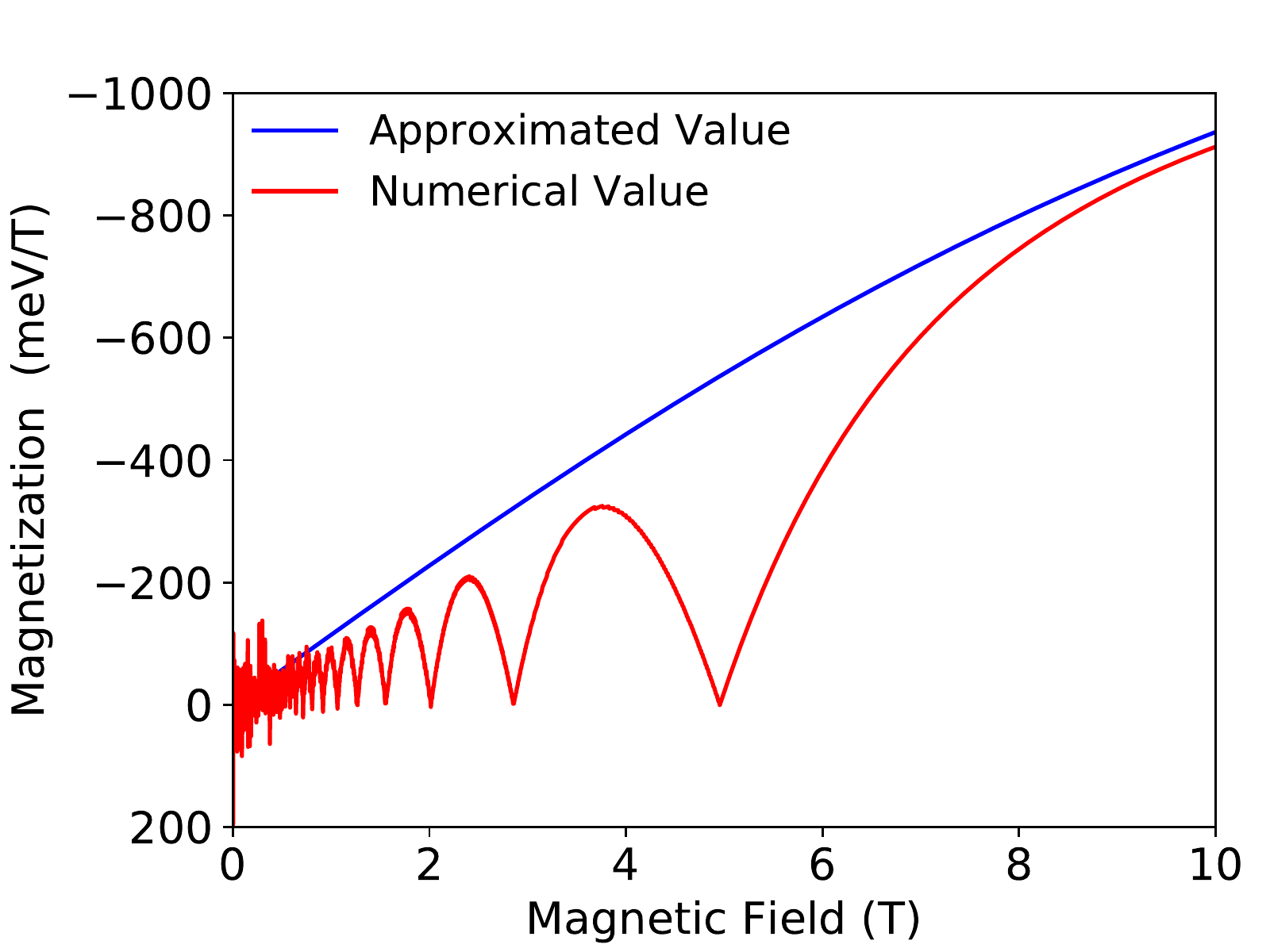}
\caption{Comparison between the approximated function for $\mathcal{M}$ and the numerical approach for a long range of magnetic field. In
the strong magnetic field regime, both approaches tend to a saturation value of $\mathcal{M}\approx-(e\hbar N^{2}_{e}/9\mu)$.}
\label{magnetization}	
\end{figure}

Expression (\ref{aqd35}) shows that the magnetization of the system is always negative across the all range of magnetic field, what indicates a diamagnetic property which should be related to a response of the system to the penetration of the magnetic field. Moreover, result (\ref{aqd35}) shows that, when $\omega_{c}\rightarrow\infty$, $\mathcal{M}=-(e\hbar N^{2}_{e}/9\mu)\approx -1606\hspace{0.05cm}\text{meV/T}$, which establishes a saturation point for magnetization. Both the numerical description from literature and the behavior curve obtained from our model are sketched in Fig. \ref{magnetization}. 

Equation (\ref{aqd34}) can be written terms of the dot area $A$ as
\begin{equation}
U(B)=\dfrac{4}{3}\dfrac{N^{2}_{e}\pi\hbar^{2}}{\mu A}\left(1+\dfrac{1}{N_{e}\left(1-\omega_{r}^{2}\right)}\right),
\label{aqd36}
\end{equation}
and the corresponding pressure reads
\begin{equation}
P(B)\approx P_{0}\left(1+\dfrac{1}{N_{e}\left(1-\omega_{r}^{2}\right)}\right).
\label{aqd37}
\end{equation}
We can check immediately that $P_{0}(B\rightarrow 0)\approx P_{0}$. Similarly to temperature in Eq. (\ref{aqd27}), result (\ref{aqd37}) reveals that the magnetic field also causes variations in the dimensions of the system.

\begin{figure}[!ht]
\centering
\includegraphics[scale=0.45]{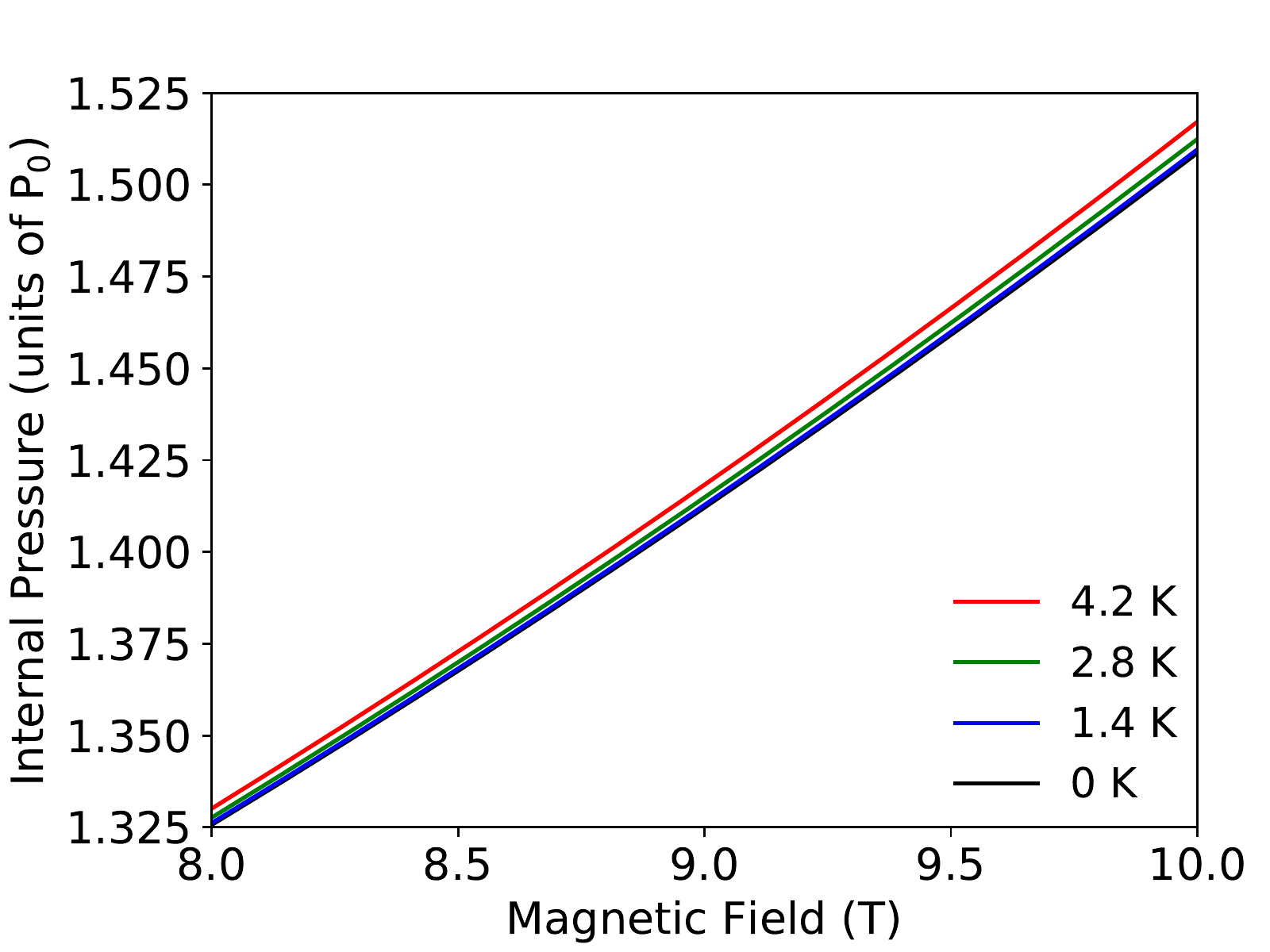}
\caption{Internal pressure (Eq. (\ref{aqd41})) as a function of the magnetic field for different values of temperature. When the temperature is increased, the effects of the magnetic field are intensified.}
\label{pressuretemp}	\end{figure}

Let us now consider both effects of temperature and magnetic field on the quantum dot simultaneously. Rewriting Eq. (\ref{aqd32}) in the form
\begin{equation}
D(\epsilon, B)=\frac{2\left(1+N_{e}\left(1-\omega_{r}^{2}\right)\right)}{\left(1-\omega_{r}^{2}\right)\epsilon}
\label{aqd38}
\end{equation}
and noting that $\omega_{r}$ does not depend on $\epsilon$, we can directly use Eq. (\ref{aqd24}) to arrive at the expression for the heat capacity due to the electrons:
\begin{equation}
C_{e}(B, T)=\dfrac{2\pi^{2}k_{B}}{3}\frac{\left(1+N_{e}\left(1-\omega_{r}^{2}\right)\right)}{\left(1-\omega_{r}^{2}\right)}\left(\frac{T}{T_{f}}\right).
\label{aqd39}
\end{equation}
Therefore, proceeding similarly as before, we obtain the expression for the internal energy
\begin{equation}
\dfrac{U(B,T)}{U(B)}=\left[1+\frac{\pi^{2}\left(1+N_{e}\left(1-\omega_{r}^{2}\right)\right)}{2N_{e}\left(1-\omega_{r}^{2}\right)}\left(\frac{T}{T_{f}}\right)^{2}\right]
\label{aqd40}
\end{equation}
and for the internal pressure
\begin{equation}
\dfrac{P(B,T)}{P(B)}=\left[1+\frac{\pi^{2}\left(1+N_{e}\left(1-\omega_{r}^{2}\right)\right)}{2N_{e}\left(1-\omega_{r}^{2}\right)}\left(\frac{T}{T_{f}}\right)^{2}\right].
\label{aqd41}
\end{equation}
The pressure profile as a function of $B$ for different values of $T$ is sketched in Fig. \ref{pressuretemp}. Although $T$ and $B$ are independent parameters, their effects on the system are connected to each other. For example, when the temperature increases, the effects of the magnetic field are intensified. Moreover, for a higher values of $B$, the influence of temperature becomes more evident.  

In summary, we have proposed a new model based on the geometric representation of a simple isosceles triangle that allows to describe the electronic and thermal properties of a quantum dot in a simpler manner. We have computed analytically some important relations of the system, such as Fermi energy, magnetization, internal energy, pressure, and heat capacity. Some of these relations have been already investigated by experimental and numerical methods. Our model predicts the existence of a force per unit length analogous to pressure in a three-dimensional thermodynamic system that increases with temperature and magnetic field strength. The most relevant results were expressed in terms of the area of the triangle, the magnetic field and the temperature. The model is also able to determine in the magnetization profile the saturation region in the strong magnetic field regime. In the profile of the internal pressure as a function of the magnetic field, we have argued that when the temperature is increased, the effects of the magnetic field are intensified. 

\section*{Acknowledgments}

This work was partially supported by the Brazilian agencies CAPES, CNPq and
FAPEMA. EOS acknowledges CNPq Grant 307203/2019-0, and
FAPEMA Grant 01852/14. This study was financed in part by the
Coordena\c{c}\~{a}o de Aperfei\c{c}oamento de Pessoal de N\'{\i}vel Superior
- Brasil (CAPES) - Finance Code 001.

\bibliographystyle{apsrev4-2}

\end{document}